\documentclass[preprint,12pt]{elsarticle}

\usepackage{amssymb}

\journal{Nuclear Instruments and Methods in Physics Research Section A}

\begin{document}

\begin{frontmatter}



\title{The Nuclear Science References (NSR) Database \\ and Web Retrieval System}

\author[label1]{B. Pritychenko}
\address[label1]{National Nuclear Data Center, Brookhaven National Laboratory, \\ Upton, NY 11973-5000, USA}
\author[label2]{E. B\v{e}t\'{a}k}
\address[label2]{Institute of Physics, Slovak Academy of Sciences, 84511 Bratislava, Slovakia}
\author[label3]{M.A. Kellett}
\address[label3]{Nuclear Data Section, International Atomic Energy Agency, \\ Vienna International Centre, P.O. Box 100, A-1400 Vienna, Austria}
\author[label4]{B. Singh}
\address[label4]{Department of Physics \& Astronomy, McMaster University, Hamilton, \\ Ontario L8S 4M1, Canada}
\author[label1]{J. Totans}

\address{}

\begin{abstract}
The Nuclear Science References (NSR) database together with its associated Web interface, is the world's only comprehensive source 
of easily accessible low- and intermediate-energy nuclear physics bibliographic information for more than 200,000 
articles since the beginning of nuclear science. 
The weekly-updated NSR database provides essential support for nuclear data evaluation, compilation and research 
activities. The principles of the database and Web application development and maintenance are described. 
Examples of nuclear structure, reaction and decay applications are specifically included. \\

\noindent The complete NSR database is freely available 
at the websites of the \linebreak National Nuclear Data Center {\it http://www.nndc.bnl.gov/nsr} 
and the \linebreak International Atomic Energy Agency {\it http://www-nds.iaea.org/nsr}.
\end{abstract}


\end{frontmatter}


\section{Introduction}
\label{sec:Intro}

The NSR database is a bibliography of nuclear physics articles, indexed according to 
content and spanning more than 100 years of research. The database originated at the Nuclear Data Project at Oak
Ridge National Laboratory as part of the systematic evaluation of
nuclear structure data \cite{ew78} and was later adopted by the 
wider research community.  It has been used since the early 1960's to produce
bibliographic citations for nuclear structure and decay data evaluations published in Nuclear Data Sheets. Periodic
additions to the database were published as ``Recent References'' issues of  Nuclear Data Sheets until 2005. 
These issues, in PDF format, are available for download from the NSR Web interface {\it http://www.nndc.bnl.gov/nsr/recref.jsp}.

In October 1980, database maintenance and updating became the responsibility of the National Nuclear Data Center (NNDC) at 
Brookhaven National Laboratory (BNL). Since then the database has been through a scope expansion, several 
modernizations, and technical improvements \cite{wi05,pr06}; however,  
the basic structure and contents have remained unchanged. In recent years, the IAEA Nuclear Data Section and 
the Nuclear Data Group at McMaster University, Canada, have joined the NSR compilation and development effort.

In this paper, we present the contents and features which make the NSR database an essential nuclear 
bibliographic source. A detailed description of the database, Web interface, and update policies are  given in the following sections.

\section{Database Scope and Structure}
\label{sec:Scope}
The NSR database aims to provide primary and secondary bibliographic information for low- and intermediate-energy nuclear physics \cite{ra96}.
Over 80 major physics journals are checked on a regular basis for relevant articles to include as primary references.
For two journals, Physical Review C and Nuclear Physics A, {\it every} article is compiled into NSR. 
Secondary references are typically conference proceedings, laboratory reports, theses, preprints, and private communications. 
Additional specific references are regularly added following the request of ENSDF evaluators, nuclear data users,  
or research centers.

The diverse contents of the database are cataloged under seven major physics topics:
\begin{center}
\begin{tabular}{p{5cm}p{5cm}}
{\sc Atomic Masses} & {\sc Nuclear Reactions} \\ [0.2cm]
{\sc Atomic Physics} & {\sc Nuclear Structure} \\ [0.2cm]
{\sc Compilation}   & {\sc Radioactivity}  \\ [0.2cm]
{\sc Nuclear Moments} \\
\end{tabular}
\end{center}

Table \ref{Table1} shows statistics pertaining to the current contents of the actual database, indicating over 200,000 articles are now compiled into NSR from over 473 different journals and many other distinct sources. 
In addition to the online database facility, {\it all} articles compiled in NSR are also preserved as hard copies in the NNDC library. 
As part of NSR services, NNDC 
can provide copies of articles from rare journals and proceedings to data evaluators and users. 

\begin{table}[hbt] 
\begin{center}
\caption[NSR Database Content as of October 2010.]{NSR Database Content as of January 2011.}
\vspace{0.2cm}
\label{Table1}
\begin{tabular}{l|r}
\hline
Database Entity & Total \\ 
\hline \hline
All references/entries & 201,848 \\
Primary references (Journal articles)	& 148,436 \\
Secondary references (Lab reports, theses, priv comm.)  & 53,412 \\
Journals ({\sc codens}) & 473 \\
Subjects & 148 \\
Authors & 86,837 \\
Nuclides & 5,002 \\
Reactions & 6,911 \\
Projectiles, ejectiles & 970 \\
\hline
\end{tabular}
\end{center}
\end{table}

\begin{figure}
\begin{center}
\includegraphics[height=7cm]{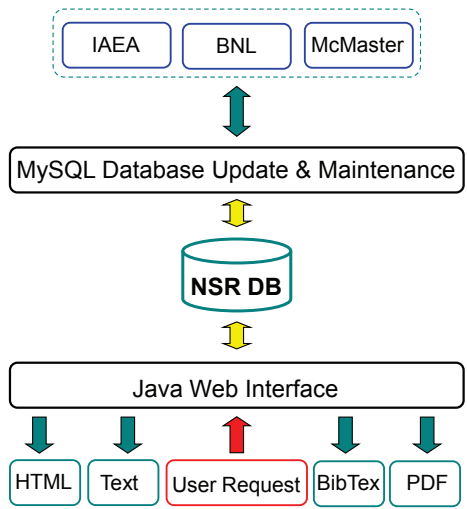}
\caption{NSR work flow.}
\label{fig1}
\end{center}
\end{figure}

The NSR database operation work flow is conducted in cooperation among three major centers: NNDC/BNL, IAEA and McMaster University. Nuclear 
physics journals are scanned for articles of interest on a weekly basis. Selected entries are compiled by one of the three 
centers and loaded into the database. NSR also enjoys a good relationship with Physical Review C authors, as they 
contribute keywords directly to the NSR database, and also the Nuclear Physics A journal, where keywords are 
prepared by the IAEA for inclusion into the database and the final manuscript prior to publishing (upon acceptance by the author). 
All of this information is processed centrally at the NNDC, where it is loaded into the master database. The complete database 
is copied to the IAEA Web server on a monthly basis. NSR database work flow is shown in Fig. \ref{fig1}. 

NSR entries include extensive information, starting with a unique eight-character identifier (NSR keynumber), 
journal/reference, publication year, article title, author list, journal digital object identifier (DOI) link, and a keyworded 
abstract (for articles reporting on appropriate physical quantities). All entries are stored in a relational database.

The current software system is based on commercially available database technologies and the Java 2 Enterprise Edition, with custom-written Java Server Pages. 
At the NNDC, this system is installed on  RedHat Linux (DELL PowerEdge 2900, 2x3.3 GHz Quad Core Xeon Processor, 64 GB RAM, 450 GB 15 kRPM hard drive) database, Web, and application servers. 
The database server is running MySQL 5 RDBMS software, while the Web server has 
the Apache 2/Tomcat 5.5/mod${\textunderscore}$jk 1.2 Web production environment installed.  Previously, the NSR database was based on Sybase ASE 15 server \cite{pr06};  
however, it was migrated to MySQL database server in 2009 during an overall hardware upgrade. 
These servers also host NNDC Web and database services. The new NSR software and hardware system is  robust and 
requires low maintenance.

\section{NSR Keywords}
\label{sec:Keywords}
The main goal of NSR is to provide bookmarks for experimental and theoretical articles in nuclear science 
using keywords. In NSR, keywords serve a dual purpose:

\begin{enumerate}
\item They are used to generate database {\it selectors}, which produce the correct article indexing and allow 
specific and detailed searches to be made quickly and easily. (Searching can also be done within the general text of entries.)
\item They allow a user to quickly determine which articles are of specific interest from a list of entries
returned following a given query.
\end{enumerate}

In preparing the keyword abstract, NSR compilers pick out the specific physical systems
being studied (isotopes, reactions, etc.), and the quantities being discussed
(i.e. measured cross sections, calculated energy levels).

By the very nature of the NSR database, the keyworded abstracts are very well structured. 
They begin with the topic identifier, as listed in section \ref{sec:Scope}, and a list of nuclei, 
nuclear reactions, or decays follows. Then the measured and/or calculated/analyzed quantities are given, 
followed by deduced (derived) quantities. 
Additional information concerning the experimental/detector details or theoretical models employed is included. 
In some cases, keywords may be provided under {\it two or more} major topics, depending on the nature of the article.

Historically, under measured quantities in NSR, we understand direct results of 
online measurements. For example, these primary quantities will 
include $\gamma$-transition energy and intensity, particle-$\gamma$ coincidences, etc. 
Other quantities, such as S-factors, log {\it ft}, and B($\lambda$) values that are often derived offline, using 
the primary data, are considered deduced quantities. 
The same philosophy applies for calculated and analyzed quantities. 

\subsection{An example explained}
\label{sec:An example explained}

In order to illustrate the use of keywords, an example is presented below in more detail.

The keyworded article includes data relevant to a series of inelastic scattering
reactions, with details of the various targets, incident/outgoing particles, and beam energies given.
Information is also included, following the specification of the reactions, on how the secondary beams
were produced. The {\it measured} quantities are then listed, followed by the {\it deduced} quantities, in this
case information relating to the levels of the scattered particles. An additional set of three sentences then 
gives detailed information relating to the nature of the findings of the work.\\

\begin{tt}
\noindent
{\small
<KEYWORDS>NUCLEAR REACTIONS $^{197}$Au($^{82}$Ge,$^{82}$Ge$'$), E=89.4 MeV/nucleon; \\
$^{197}$Au($^{84}$Se,$^{84}$Se$'$), E=95.4 MeV/nucleon; $^{9}$Be($^{82}$Ge,$^{82}$Ge$'$), E=87.6 MeV/ \\
nucleon; $^{9}$Be($^{84}$Se,$^{84}$Se$'$), E=92 MeV/nucleon, [$^{82}$Ge and $^{84}$Se secondary \\ 
beams from $^{9}$Be($^{86}$Kr,X), E=140 MeV/nucleon]; measured E$_{\gamma}$, I$_{\gamma}$, $\sigma$, \\
(particle)$\gamma$-coin; $^{82}$Ge, $^{84}$Se; deduced levels, J, B(E2), T$_{1/2}$. \\
Intermediate energy Coulomb excitation and inelastic scattering. \\
Comparison with systematics of B(E2) values for first 2$^{+}$ state \\
in N=50 isotones for Z(even)=30-42 and even-even Ge (A=64-82) and \\
Se (A=68-84) isotopes, and with shell-model calculations. \\
Systematics of first 3$^{-}$ states in even-even Se (A=74-82) and N=50 \\
isotones. \\
}
\end{tt}

\subsection{Keyword preparation}
\label{sec:Keyword preparation}

Keyword abstract preparation is the  part of  NSR work flow  that is difficult to automatize. 
All abstracts are prepared manually, which ensures
the high quality of the keywords. However, this limits the volume of processed articles. To overcome 
this problem, NNDC started to work with XSB, Inc. in 2010 on ``Semantic Analysis of Nuclear Physics Publications and Automatic NSR Keyword Generation''. 
 Selected articles in PDF format are converted to text and analyzed for keywords using the Apache UIMA Framework \cite{uima}. 
The initial implementation and testing of semantic analysis is showing promise in the partial-automation of NSR keyword abstract preparation and 
the consequent reduction of required manpower. However, human effort, albeit on a reduced scale, will always be necessary for ensuring the 
quality of any automatically generated keyword abstracts.

\subsection{Added value}
\label{sec:Added value}

In the last two decades, bibliographic databases have faced increasing competition from Web search engines and corresponding databases. 
Google and its Advanced Scholar Search ({\it http://scholar.google.com/}) complement NSR and provide users with extensive lists of 
indexed information that match the search criteria ordered by frequency of use.

What differentiates NSR from these essentially text pattern-matching search engines is the level of 
detail provided in the keyword abstracts and the resulting indexes which direct the user's search. 
For example, a Web search engine cannot differentiate between $^{32}$Mg and 32mg, nor $^{31}$Na and 31na, 
since these are equivalent in plain text. Hence a Web search engine would return over 591,000  prescription drug links containing 32mg (milli-grams) 
of a particular product, and in 338,000 cases, $^{31}$Na would be confused with some electrical leakage current of 31 nano-amperes.
NSR, as a dedicated nuclear physics database, will produce a clean output containing only the 175 and 111 relevant articles for $^{32}$Mg and $^{31}$Na, respectively.

\section{NSR Retrievals}
\label{sec:Retrievals}
\begin{figure}
\begin{center}
\fbox{\includegraphics[height=10cm]{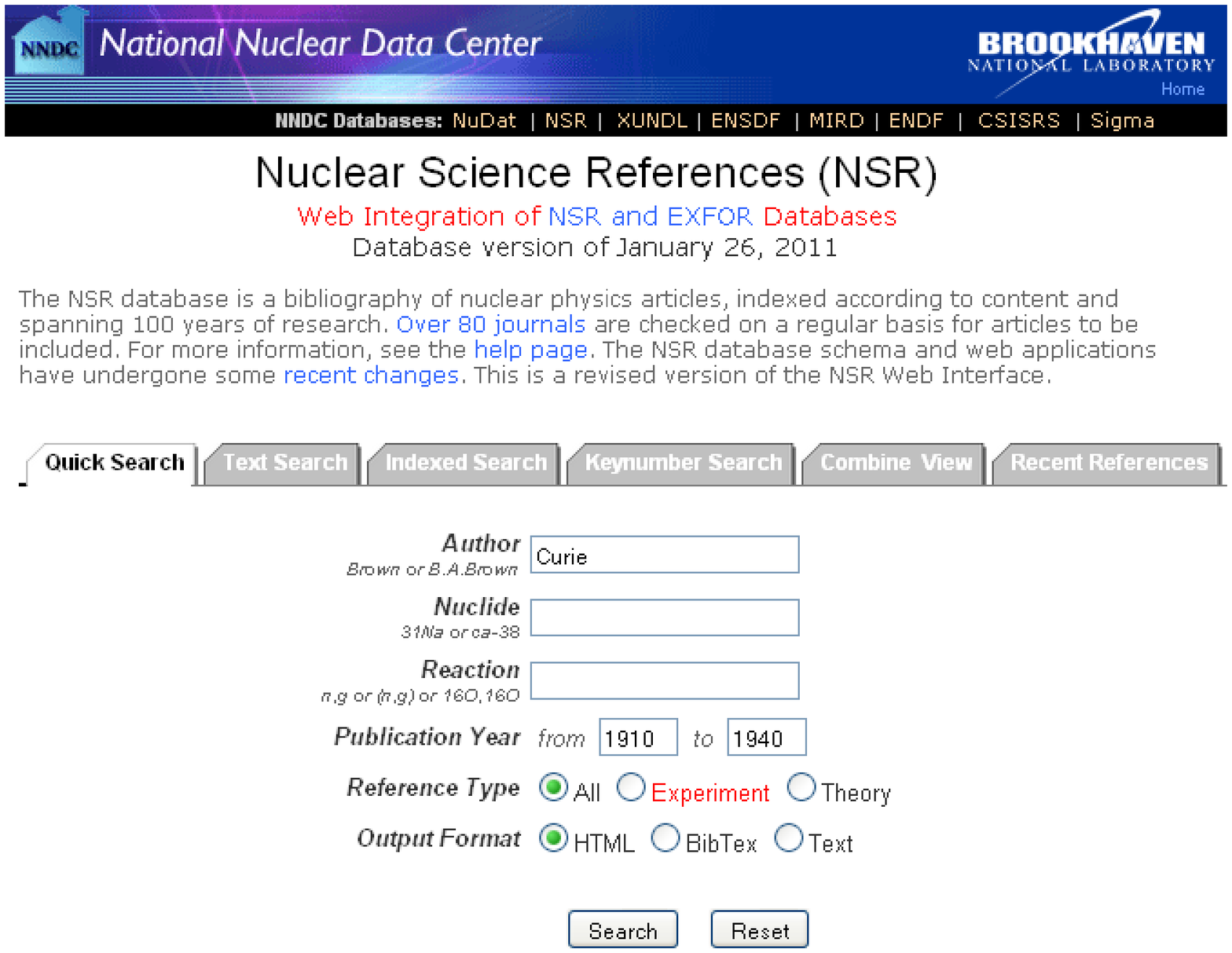}}
\caption{NSR Web Interface {\it http://www.nndc.bnl.gov/nsr}. Example of Boolean search for author Curie and 1910-1940 time range.}
\label{fig3}
\end{center}
\end{figure}

The NSR Web Retrieval Interface is an integral part of both the NNDC and IAEA Web Services \cite{pr06,ph10}. 
The Web interface is based on current Java technologies and provides retrievals of the database content in HTML, Text, BibTex, and PDF formats. 
As shown in Fig. \ref{fig3}, the main Web interface actually consists of six sub-interfaces:

\begin{center}
\begin{tabular}{l}
Quick Search \\ [0.2cm]
Text Search \\ [0.2cm]
Indexed Search \\ [0.2cm]
Keynumber Search \\ [0.2cm]
Combine View \\ [0.2cm]
Recent References \\ [0.2cm]
\end{tabular}
\end{center}

The Quick Search allows a quick look-up of references for a given author, nuclide, or reaction within a publication period. 
The Text Search allows plain text searching of the title and keyword fields, whilst an Indexed Search allows a 
Boolean {\sc and} search over several indexed categories (e.g. author, nuclide, etc.). 
The Keynumber Search retrieves the information for a specific article(s) given the NSR keynumber(s).
This type of specific retrieval is in large demand by nuclear structure evaluators. 
Finally, Combine View provides analysis and combination opportunities for previous retrievals, 
whilst Recent References provides downloads of quarterly compilation collections in PDF format. 
Entries for certain theory and review articles may not include keyword abstracts, if they do not deal 
with specific nuclides or reactions. These may still be retrieved in other ways, including author, text, or keynumber searches. 

\subsection{Simple NSR retrievals}

NSR retrievals are generally based on searching the indexed quantities created from the keywords but can also be made
simply by specifying an author and/or publication year. For example, searching on author {\it Curie} between {\it 1910} and {\it 1940} results
in three entries, one of which is shown here: \\

\begin{tt}
\noindent
{\small
1931CU01 \\
Revs.Modern Phys. 3, 427 (1931) \\
M.Curie, A.Debierne, A.S.Eve, H.Geiger, O.Hahn, S.C.Lind, S.Meyer, \\
E.Rutherford, E.Schweidler \\
The Radioactive Constants as of 1930 \\
doi: 10.1103/RevModPhys.3.427 \\
}
\end{tt}

As is shown, each entry stored in the database corresponds to a full bibliographic reference that is uniquely 
identified by an eight character alphanumeric keynumber, e.g. {\it 1931CU01}.
The first four digits of the keynumber give the publication year of the corresponding reference, 
followed by the first two characters of the first author's surname, and finally a two digit incremental sequence number allocated during the database loading. 

In this particular case, no keywords have been created, but one can clearly see the full bibliographic reference 
to volume 3 of the journal {\it Reviews of Modern Physics}, page 427, followed by the full list of authors, 
the paper title, and finally the DOI (Digital Object Identifier),
which through the Web retrieval system provides a direct link to the original publication webpage, although full 
access to the paper may be subscription dependent.

Most of the entries ($\geq$80$\%$) include keyword abstracts, which provide a brief summary of the subject matter 
in the given reference. As already mentioned, these are used to generate the indexed quantities in the database. 

In the example below, the complete entry {\it 2000SI01} is shown using the {\sc Compilation} topic. 
The keywords show a list of all nuclei, of {\it mass 163}, reported in this article: \\

\begin{tt}
\noindent
{\small
2000SI01 \\
Nucl.Data Sheets 89, 1 (2000) \\
B.Singh, A.R.Farhan \\
Nuclear Data Sheets for A = 163 \\
COMPILATION $^{163}$Gd, $^{163}$Tb, $^{163}$Dy, $^{163}$Ho, $^{163}$Er, $^{163}$Tm, $^{163}$Yb, $^{163}$Lu, $^{163}$Hf, \\
$^{163}$Ta, $^{163}$W, $^{163}$Re, $^{163}$Os; compiled, evaluated structure data. \\
doi: 10.1006/ndsh.2000.0001 \\
}
\end{tt}

A partial example below illustrates the keywords for a reference using the {\sc Nuclear Structure} ``topic'': \\

\begin{tt}
\noindent
{\small
NUCLEAR STRUCTURE $^{192,193,195,199,200,201}$Pb; analyzed magnetic rotational \\
bands signature splitting. Tilted-axis cranking, particle-rotor \\
models. \\
}
\end{tt}

\subsection{Advanced NSR retrievals}
There are two major types of NSR retrievals that are often ignored by new users, i.e. Indexed Search and Text Search.

The Indexed Search uses the database {\it selectors} and allows a highly targeted Boolean {\sc and} search of up 
to three specific parameters.
The full list of available parameters is shown in Table \ref{Table2}. 
As can be seen, all relevant fields for an entry can be searched. Although each search only allows three parameters
to be specified, the results of multiple searches can be cross-compared using the Combine View subinterface, where
the Boolean {\sc or} and {\sc not} operators can be specified, meaning that selective searches can be carried out.

\begin{table}
\begin{center}
\caption{NSR Indexed Search.}
\vspace{0.2cm}
\label{Table2}
\begin{tabular}{l|p{8.5cm}}
\hline
Indexed Quantity & Description  \\
\hline \hline
Author & Author's last name, and (optionally) one or two initials, e.g. B.A. Brown. \\ [0.2cm]
FirstAuthor & References where given author is {\it first} in the \newline author list. \\ [0.2cm]
Nuclide & Nuclide of interest in the format BR-76 or 76BR.  \\ [0.2cm]
Target, Parent, and Daughter & Same format as for ``Nuclide''. \\ [0.2cm]
Reaction & Reaction of interest, e.g. (p,a) or n,p.  \\ [0.2cm]
Incident & Incident beam particle, e.g. n or p.  \\ [0.2cm]
Outgoing & Outgoing particle in a reaction, e.g. p or a. \\ [0.2cm]
Subject  & ``Measured'', ``Deduced'', ``Calculated'' required at beginning of keyword phrase, 
followed by an indexed quantity, e.g. $\sigma$, T$_{1/2}$, B($\lambda$), Q-values, S-factors.  \\ [0.2cm]
Journal ({\sc coden}) & Five-letter code associated with a given journal, e.g. \newline {\sc prvca} - Phys.Rev. C, 
\newline {\sc nupab} - Nucl.Phys. A, \newline {\sc nimae} - Nucl.Instr.Meth. A  \\ [0.2cm]
Topic & {\sc Nuclear Structure}, \newline {\sc Nuclear Reactions}, \newline {\sc Radioactivity}  \\ [0.2cm]
Z(range),A(range) & Numerical value for Z or A, and/or range.\\
\hline
\end{tabular}
\end{center}
\end{table}

The Text Search sub-interface can be used to search for references based on text matches in the title and keyword fields. 
Searches can be restricted to one of these fields, allowing searches of articles even without keyword abstracts, or can span both. 

\subsection{NSR retrieval statistics}
\begin{figure}
\begin{center}
\includegraphics[height=7cm]{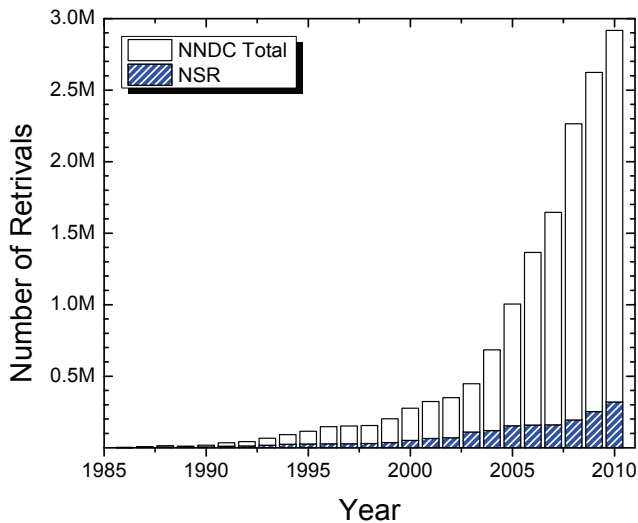}
\caption{The time evolution of the number of electronic retrievals - web, telnet and FTP - from 1986 to 2010.}
\label{fig4}
\end{center}
\end{figure}

An important part of monitoring NSR operation is a correct estimate of the database usage. NSR retrieval statistics are very conservative \cite{tu11} 
and completely based on a count of successful database retrievals - any Web browser hits are ignored. Retrievals for non-existing entities in the 
database (e.g. author, nuclide) which produce an empty output file are consequently not counted. Complete information for 
each retrieval is recorded in a separate statistics database. Normally, there are $\sim$700 retrievals/day with $\sim$800 references/retrieval 
from the NNDC database alone. 
Such a large number of references/retrievals is a direct result of the broad selection criteria entered into the Quick Search subinterface 
and the database size. A complementary way of estimating NSR statistics  is based on analysis of the Apache Web Server log. These logs, after removal of crawler and/or search engine activity, produce comparable results
to the database retrievals count method. The time evolution for NSR retrievals at NNDC over the last 25 years is shown in Fig. \ref{fig4}. 

\begin{figure}
\begin{center}
\includegraphics[height=7cm]{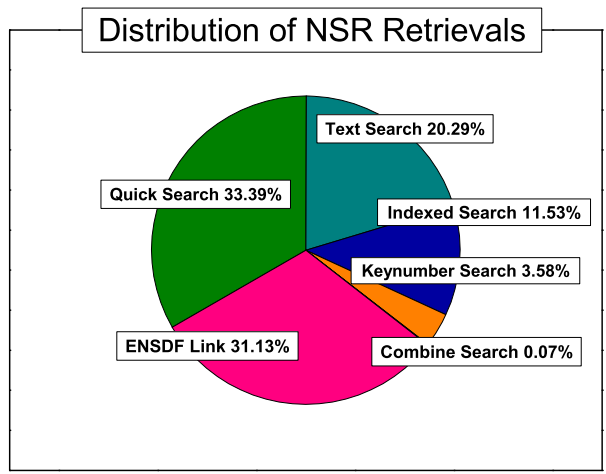}
\caption{Distribution of NSR retrievals for different types of search activities.}
\label{fig5}
\end{center}
\end{figure}

Further analysis of NSR database retrievals reveals information on patterns of NSR usage, users Web browsers and operating systems, separate users and organizational 
activities over time, etc. These patterns are used for NSR database and Web interface quality assurance purposes. 
The distribution of NSR retrievals for different types of search activities is shown in Fig. \ref{fig5}. A large volume of 
ENSDF database connections (ENSDF Link) is due to direct access from the ENSDF, B(E2) and $\beta$$\beta$-decay webpages. 
More information on this subject is provided in the next section.

\section{NSR Applications} 
\label{sec:Applications}
\begin{figure}
\begin{center}
\includegraphics[height=7cm]{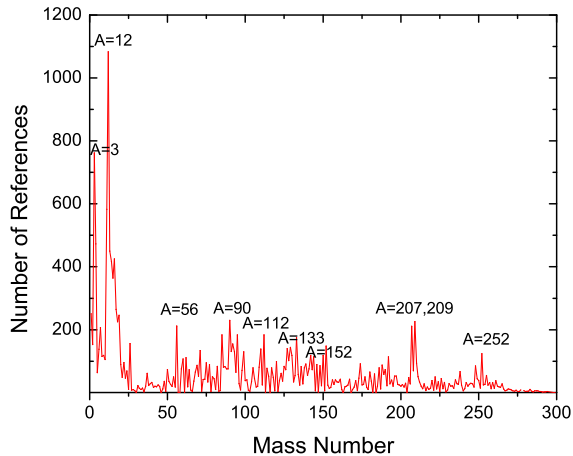}
\caption{Mass number distribution of references unevaluated by the NSDD network, as of October 2010.}
\label{fig6}
\end{center}
\end{figure}

The NSR database was initially created to support the Evaluated Nuclear Structure Data File (ENSDF) \cite{ensdf,bu90,tu96}. All references in ENSDF 
evaluations are specified by their NSR keynumbers.  Regular NSR database 
updates serve as an indicator for the international Network of Nuclear Structure and Decay Data Evaluators (NSDD) \cite{nsdd} on the requirement to
revisit a particular isobaric mass chain. If a large number of new experimental studies for a particular nuclide or mass chain have been added since the last ENSDF 
evaluation update, this would immediately alert the NSDD network about the need for a  re-evaluation. 
Fig. \ref{fig6} shows number of references as a function of mass number, which has not yet been included in the ENSDF database. 
The large number of new articles for light nuclei is due to the general 
availability of these projectiles in rare isotope beam research.

In addition to ENSDF mass chain or vertical evaluations \cite{ensdf,fi96}, NSR is actively used in a large number of horizontal 
evaluations of atomic masses \cite{am1,am2}, B(E2) values \cite{ra01,pr11}, and magnetic moments \cite{st05}. 
The NSR database and Web interface are also linked to a large number of other nuclear databases: ENSDF, XUNDL, and EXFOR \cite{ensdf,xundl,exfor}. 
Thus when a particular reference forms part of a compilation or evaluation in one of these other databases, NSR will provide a direct Web 
link to the publication.

Finally, the NSR database has a broad usage spectrum worldwide, with more than half of NSR retrievals coming from users in research and educational fields. 
Web users from many research centers and universities such as MSU/FRIB, Yale, FSU, GSI, JINR, and RIKEN actively use NSR for data-mining purposes and interact with the database staff. 
NSR user feedback is always welcome, as this can help in defining areas of future improvement.

\section{Conclusion and Outlook}
\label{sec:Conclusions}
The NSR database and its Web interface are available \linebreak through both the NNDC ({\it http://www.nndc.bnl.gov/nsr}) and IAEA \linebreak ({\it http://www-nds.iaea.org/nsr}) websites, 
and provide transparent and easy access to nuclear 
physics bibliographic information with direct links to the original articles and data provided, where possible. 
Recent additions include many features for nuclear scientists and, specifically, {\it reaction} data users, such as user-friendly Web retrievals, Web 
integration with the EXFOR database and improvements in NSR terminology/keywording. NSR has much potential in modern physics, 
as it is the major nuclear database that allows searches for rare isotope beam reactions. 

Further implementation of the latest technologies and computer-aided keywording and semantic procedures will improve database maintenance 
procedures, reduce the workload required for keyword abstract preparation, and ensure the continuing high quality of database content. 
Exponential growth of electronic access to nuclear data \cite{pr06} requires continuing effort to satisfy current needs and future demands.

\section{Acknowledgments}
\label{sec:Acknowledgements}
We are grateful to M. Herman (BNL) and D. Abriola (IAEA) for their constant support of this project, to  D.F. Winchell (XSB, Inc.) for significant technical contributions,  
to J. Choquette (McMaster University) for useful suggestions, and to M. Blennau (BNL) and V. Unferth (Viterbo University) for a careful reading of the manuscript.  
This work was sponsored in part by the Office of Nuclear Physics, Office of Science of the U.S. 
Department of Energy under Contract No. DE-AC02-98CH10886 with Brookhaven Science Associates, LLC.





\bibliographystyle{model1a-num-names}
\bibliography{<your-bib-database>}



\end{document}